\documentclass[12pt,preprint]{aastex}

\shorttitle{GRBs in Relativistic Turbulence Model}
\shortauthors{Lin et al.}

\begin{document}

\title{Revisiting the Light Curves of Gamma-ray Bursts \\
in the Relativistic Turbulence Model}

\author{Da-Bin Lin\altaffilmark{1,2},
Wei-Min Gu\altaffilmark{1},
Shu-Jin Hou\altaffilmark{1},
Tong Liu\altaffilmark{1},
Mou-Yuan Sun\altaffilmark{1},
and Ju-Fu Lu\altaffilmark{1}}

\altaffiltext{1}{Department of Astronomy and Institute of Theoretical Physics and Astrophysics, 
Xiamen University, Xiamen, Fujian 361005, China;
dabinlin@xmu.edu.cn, lujf@xmu.edu.cn}
\altaffiltext{2}{Department of Physics and GXU-NAOC Center for Astrophysics and Space Sciences,
Guangxi University, Nanning 530004, China}

\begin{abstract}
Rapid temporal variability has been widely observed in the light curves of
gamma-ray bursts (GRBs). One possible mechanism for such variability
is related to the relativistic eddies in the jet.
In this paper, we include the contribution of the
inter-eddy medium together with the eddies to the gamma-ray emission.
We show that the gamma-ray emission can either lead or lag behind the observed synchrotron
emission, where the latter originates in the inter-eddy medium
and provides most of seed photons for producing gamma-ray emission through the inverse-Compton scattering.
As a consequence, we argue that the lead/lag found in non-stationary
short-lived light curves may not reveal the intrinsic lead/lag of
different emission components.
In addition, our results may explain the lead of gamma-ray emission
with respect to optical emission observed in GRB 080319B.
\end{abstract}


\section{Introduction} \label{sec:Introduction}
Gamma-ray bursts (GRBs) are the most extreme explosive events in the Universe.
They are known to be highly variable, and the temporal structure exhibits
diverse morphologies (\citealp{Fishman1995}), which can vary from
a single smooth large pulse to extremely complex light curves
with many erratic short pulses.
It is believed that such kind of variability, especially fast variability,
may provide an interesting clue to the nature of GRBs
(e.g., \citealp{Morsony2010}).

The millisecond variabilities observed in the prompt phase have
led to the development of the internal shock model
(\citealp{Piran1993}; \citealp{Katz1994}; \citealp{Rees1994}).
In this model, an ultrarelativistic jet is ejected with a fluctuating velocity profile.
When a fast latter ejected portion of the jet catches up with the
slow earlier ejected one, a pair of internal shocks is formed.
Each pulse in the light curve of bursts corresponds to one such collision
(\citealp{Kobayashi1997}; \citealp{Maxham2009}).
This is the reason for many erratic short pulses in the light curves.
The internal shock model was discussed and simulated numerically in a number of works.
It was found that only a small fraction of the total kinetic energy can be dissipated in this process
(\citealp{Kobayashi1997}; \citealp{Daigne1998}).
However, a detailed study suggests that the radiative efficiency
of some GRBs can be up to 90\% (\citealp{Zhang2007}), which is
difficult to be produced within the straightforward internal shock model.

In the internal shock model, the variability of prompt emission is attributed to
the history of central engine activity. However, the observed variability may originate in the emission region.
This scenario requires that the emission region should not be uniform.
The external shocks form while the outflow is slowed down by significantly small circumburst clumps
can produce highly variable light curves (e.g., \citealp{Dermer1999}).
However, this process is inefficient (\citealp{Sari1997}; see \citealp{Narayan2009} for details).
Alternatively, Lorentz-boosted emission units in the jet,
such as mini-jets (\citealp{Lyutikov2003}; \citealp{Giannios2009}) or relativistic
turbulent eddies (\citealp{Narayan2009}),
can also produce strong variability of gamma-ray emission.
According to this scenario and with the consideration of gamma-ray emission
only from eddies,
\cite{Lazar2009} reproduced the fast variability observed in a sub-sample of GRBs.
However, the amplitude of subjacent smooth component in the
simulated light curve is too low compared with observations
(e.g., Figure~2 of \citealp{Lu2012}),
where the light curve is roughly divided into two components:
a smooth component underneath the light curve and a rapid variability
that is superimposed on the smooth component
(e.g., \citealp{Gao2012}; \citealp{Dichiara2013} and references therein).
It should be noted that the light curves in the realistic situation are more complex,
and the relativistic turbulence model for fast variability is only
applicable to a sub-sample of GRBs, such as GRBs with erratic symmetric short pulses (\citealp{Lazar2009}).
On the other hand, \citet{Kumar2009} showed that the inter-eddy medium
is predominant in the observed synchrotron emission and makes significant
contribution to the gamma-ray emission, which is produced by inverse-Compton (IC)
scattering the synchrotron photons from eddies and inter-eddies.
The purpose of this work is to revisit the relativistic turbulence model
by including the role of inter-eddy medium in producing the light curve.

The paper is organized as follows.
The relativistic turbulence and the kinematic toy model for jet radiation
are described in Section~2.
The simulated light curves for different conditions are
presented in Section~3, in which
the lead/lag of gamma-ray emission with respect to the observed synchrotron emission
is our main focus.
Conclusions and discussion are made in Section~4.

\section{Relativistic turbulence model} \label{sec:Relativistic_Turbulence model}

The relativistic turbulence model is well described in the work of
\cite{Narayan2009}, \cite{Lazar2009}, and \cite{Kumar2009}.
In this section, we take a brief description of this model following the above works.
More details can refer to these papers.

We first introduce three frames related to the relativistic turbulence model:
the lab frame; the shell's frame,
denoted by a prime, which is boosted radially with a Lorentz factor
$\Gamma$ relative to the lab frame; and the frame of an eddy, denoted
by two primes, which is boosted by $\gamma'_t$ relative to the shell frame.
In the relativistic turbulence model, the fluid in the jet consists of eddies and an inter-eddy medium.
The eddies are considered with a typical Lorentz factor $\gamma'_t$ in the shell frame
and a typical size $l''_t \sim R/(\gamma'_t\Gamma)$ in their respective frame,
where $R$ is the distance of emission region to the jet base.
The filling factor of eddies in the jet is described with a parameter $f$,
and thus around $f{\gamma'}_t^3$ eddies can be found in a causally connected volume ($\sim R^3/\Gamma^3$).
Owing to the collision with other eddies, an eddy is not
likely to travel along a perfectly straight line.
In order to describe this behavior, $\tau'=R/(\gamma'\Gamma c)$ is introduced,
which corresponds to the time for the change of eddies
velocity orientation with an angle of $1/\gamma'$.
For the inter-eddy medium, it can be discretized into ``inter-eddy'',
which has the same size and is associated with a Lorentz factor ${\gamma'}_{it}=1$ in the shell frame.

The kinematic toy model for jet radiation considers
a shell which is divided into discrete randomly distributed emitters (eddies or inter-eddies),
and the emitters radiate as the shell moves from $R_0$ to $2R_0$.
During this period, the movement direction and position of eddies continuously change.
In order to model this dynamic process,
a set of successive shells between $R_0$ and $2R_0$ are introduced.
Each new shell is constructed with randomly distributed emitters,
representing the change in movement direction and position of eddies.
The time difference between two shells is $\tau=\tau'\Gamma$,
and the thickness $\Delta$ of shells is described with a parameter $d( \geqslant 1)$, i.e., $\Delta=d R/{\Gamma ^2}$.
The thickness due to the intrinsic expansion of shell is $\Delta \sim R/{\Gamma^2}$, i.e., $d=1$.
Then, the situation of $d > 1$ reveals that the width of shell
is determined by the duration of the central engine activity.
In the present work, we study the case of $d=1$
and the emitters are described with its center position rather than its filling region.
The Doppler shift from an emitter is:
\begin{equation}
\Lambda =[{\gamma(1-\frac{\upsilon}{c}\cos\alpha)}]^{-1}  ,
\label{eq: Doppler shift}
\end{equation}
where $\gamma$, $\upsilon=c\sqrt{1-1/\gamma^2}$, and $\alpha$ are the Lorentz factor, velocity, and the angle
between the emitter velocity and the line to the observer (both in the lab frame).
If the emitter velocity ${\vec{\upsilon'}}$ is in the direction
with a polar angle $\theta'$ and an azimuthal angle $\phi'$
relative to the radial direction in the shell frame,
the emitter would move with a Lorentz factor
\begin{equation}\label{eq: Lorentz_Transform}
{\gamma} = \Gamma {\gamma'}\left( {1 + {\upsilon _{\rm j}\upsilon' \over c^2}\cos \theta'} \right)
\end{equation}
in the lab frame, and the polar angle $\theta$ and azimuthal angle $\phi$
relative to the radial direction in the lab frame satisfy
\begin{equation}\label{eq: Angle_Transform}
\tan \theta  = \frac{{{\upsilon'}\sin \theta '}}{{{\Gamma}
\left( {{\upsilon _{\rm{j}}} + {\upsilon'}\cos \theta '} \right)}},
\end{equation}
\begin{equation}
\phi=\phi',
\end{equation}
where ${\upsilon _{\rm{j}}}$ is the velocity of jet.
Then it is easy to find the relationship:
\begin{equation}
\cos\alpha=-\sin\theta\cos\phi\sin\theta_j+\cos\theta\cos\theta_j,
\end{equation}
where $\theta_j$ is the latitude of the emitter in the shell
observed in the lab frame.

The radiation mechanism for gamma-ray emission observed in the prompt
phase is the IC scattering of the synchrotron photons from eddies and inter-eddies.
Since the seed photon field, i.e., synchrotron photons from eddies and inter-eddies,
is the same for IC scattering of eddies and inter-eddies,
the observed peak frequency of the IC spectrum
from eddies should be close to that from inter-eddies
based on the relation of ${\gamma''}_e{\gamma'}_t={\gamma'}_e$.
Here, ${\gamma''}_e$ (${\gamma'}_e$) is the thermal Lorentz factor of electrons in the eddies (inter-eddies)
and the seed photon field is roughly isotropic and homogeneous in the shell's frame.
Therefore, the gamma-ray emission to the observer from an eddy or an inter-eddy can be described as
(see equations 52 and 53 in \citealp{Kumar2009}):
\begin{equation}\label{eq: IC_Flux}
F_{\rm IC} \propto \sigma_T(n_e \gamma^3)f'_{\rm syn}  \Lambda^{3} ,
\label{eq: Energy}
\end{equation}
where $f'_{\rm syn}$ is the synchrotron flux as seen by a typical electron in the inter-eddy medium,
$n_e$ is the number of electrons in an emitter.
For the relativistic turbulence model, the parameters
$n_{e,t}$, $n_{e, it}$, and $n_{it}$ should satisfy the relation of $n_{e, it}n_{it}=n_{e,t}(1-f){\gamma_t'}^3$,
where $n_{e,t}$ ($n_{e, it}$) is the number of electrons in an eddy (inter-eddy)
and $n_{it}$ is the number of inter-eddy in the inter-eddy medium.
In Equation~(\ref{eq: Energy}), we adopt the assumption in \citet{Kumar2009}
that, at a fixed observer time, the observer receives radiation
from only a fraction ($\sim 1/\gamma'_t$) of the electrons in the eddy
owing to the time dependence of eddy velocity direction.
Following the spirit of \cite{Lazar2009},
we assume $f'_{\rm syn} \propto 1/(\psi R)^{\alpha}$ and $\alpha=1$.
We also examine the light curves of gamma-ray emission with other values
of $\alpha$, which plays negligible effects on the profile of light curves
and on the lead/lag between gamma-ray emission and observed synchrotron emission.
For simplicity, the pulse produced by a single eddy is assumed as a Gaussian profile (e.g., \citealp{Lazar2009}; \citealp{Maxham2009})
\begin{equation} \label{eq:Peak-Duration}
F(t)=F_{\rm IC} \exp \left[ { - {{{{\left( {t - {t_p}} \right)}^2}} \over {2{{\left( {\delta t} \right)}^2}}}} \right] ,
\end{equation}
where $t_p$ is the time of peak flux, and $\delta t$ takes the form (\citealp{Lazar2009}):
\begin{equation} \label{eq:Peak-Duration}
\delta t \approx {R\psi}/(\gamma'\Gamma c).
\end{equation}
\cite{Norris1996} showed that pulses in some GRBs rise more quickly than they decay.
This is different from the short pulses produced by the eddies,
which may be statistically symmetric (\citealp{Lazar2009}).
However, the causes of variability in the GRB light curves may be diverse (\citealp{Gao2012}),
and therefore the shape of pulses may be different for different GRB.
The evidences, i.e., symmetric short pulses, can be found in the Appendix of \cite{Norris1996}.
It should also be noted that the IC radiation mechanism discussed above typically predicts $R_{\rm IC} \gtrsim 1$
(\citealp{Kumar2009}; \citealp{Beniamini2011}; \citealp{Guetta2011}),
where $R_{\rm IC}$ is the ratio of the fluence in the second-order IC component
to the fluence in the first-order IC component (i.e., gamma-ray emission discussed in the present work).
However, the observations of most GRBs do not support this behavior, i.e., $R_{\rm IC} \gtrsim 1$
(\citealp{Beniamini2011}; \citealp{Guetta2011}; \citealp{Ackermann2012}; \citealp{Ackermann2013}).
Then, overestimating the value of $R_{\rm IC}$ may be an issue in the relativistic turbulence model in its current stage.

The light curves of gamma-ray emission discussed below are produced based on
the kinematic toy model of jet radiation and Equations (\ref{eq: IC_Flux})-(\ref{eq:Peak-Duration}).
In the simulations, the inter-eddies are uniformly distributed in the jet shell
and a significantly large value of $n_{it}$ is chosen
in order to produce a smooth light curve of gamma-ray emission from inter-eddies.
According to the relativistic turbulence model, the parameters $\gamma'_t$, $f$,
and the distribution of eddy orientation will determine the main properties
of light curves.
Since the typical value of ${\cal V}\equiv t_{\rm burst}/t_{\rm var}=\gamma'^{2}_t$ is around $100$ (\citealp{Narayan2009}),
we adopt $\gamma'_t=10$ in the present work.

\section{Numerical simulations for the light curves}

We present the simulated light curves with consideration of the emission from inter-eddies in this section,
and focus on the lead/lag of gamma-ray emission with respect to the observed synchrotron emission.
In the realistic situations, the movement direction of eddies continuously changes with time,
and may be concentrated in some directions.
Then, we model this behavior with different $|\mu'|$ conditions, i.e., $|\mu'|\leqslant a (\geqslant a)$,
which is used in our simulations.
Here, $\mu'=\cos\theta'$, $0\leqslant a \leqslant 1$, and $|\mu'|\leqslant a (\geqslant a)$ means that
the value of $\mu'$ for an arbitrary eddy varies with time and is randomly taken from $[-a,a]$ ($[-1,-a]\cup[a,1]$).
As shown below, the different $|\mu'|$ conditions may result in different lead/lag
of gamma-ray emission with respect to the observed synchrotron emission.

\begin{figure}
\plotone{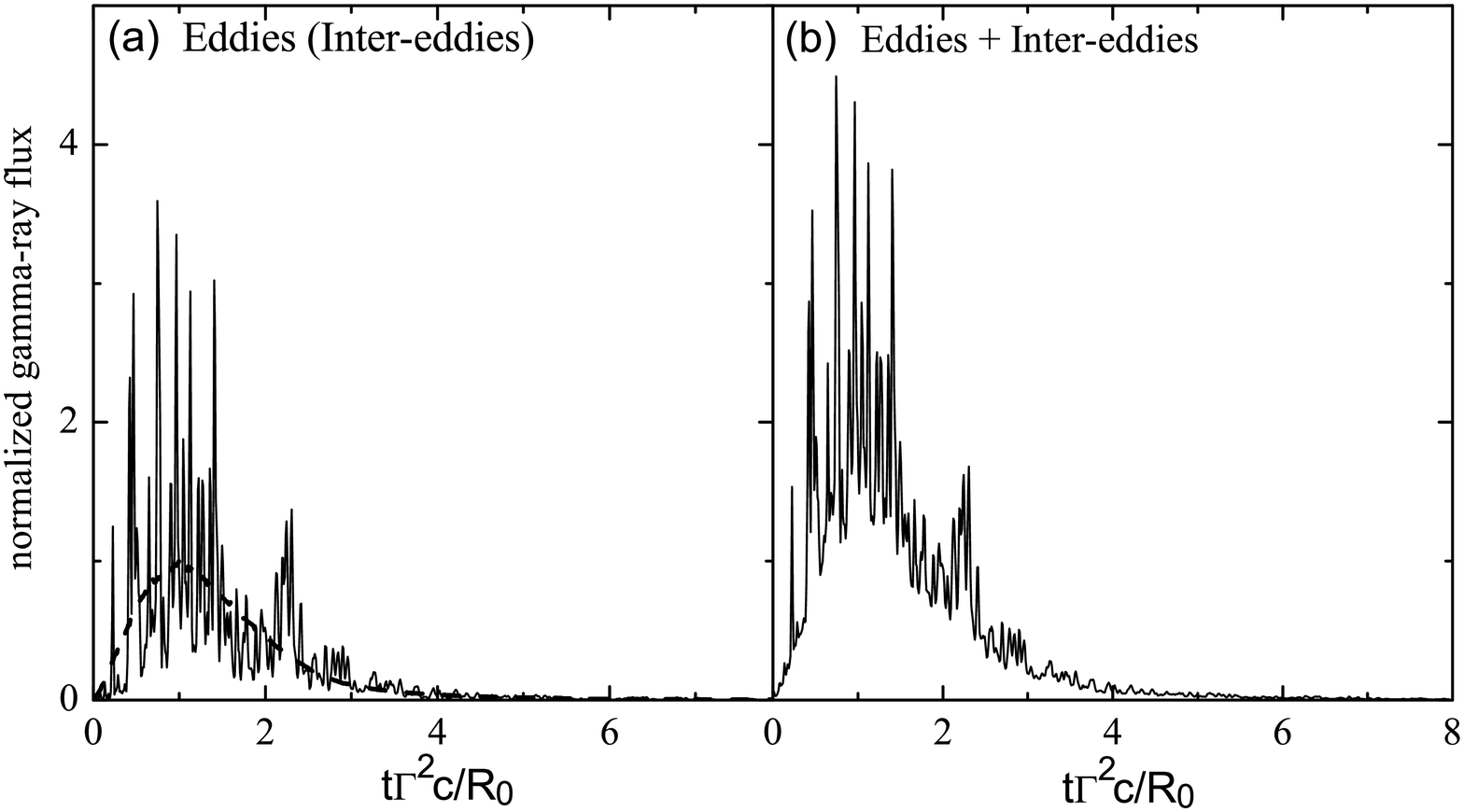}
\caption{Simulated light curves of gamma-ray emission from the eddies
(solid curve in the left panel), from the inter-eddies
(thick dashed line in the left panel),
and from both the eddies and the inter-eddies (the right panel),
with the filling factor $f=0.8$.}
\label{F:Lightcurve_Comparison}
\end{figure}

Figure~\ref{F:Lightcurve_Comparison} depicts the simulated light curves of gamma-ray emission
from the eddies (solid curve in the left panel), from the inter-eddies (thick dashed line in the left panel),
and from both the eddies and the inter-eddies (the right panel).
In this figure, the filling factor $f$ is $0.8$,
the orientation of eddy velocity is isotropic in the shell frame
(i.e., $|\mu'|\geqslant 0$ or $|\mu'|\leqslant 1$),
and the value of vertical axis is the gamma-ray flux,
which corresponds to the gamma-ray emission around the peak frequency of IC spectrum
and is normalized with the peak flux of that from inter-eddies.
Similar to the results presented in the work of \cite{Lazar2009},
the light curve produced only by eddies can be depicted as a FRED (``fast rise exponential decay''),
which looks like many observations.
However, the amplitude of subjacent smooth component is too low
compared with that of observations, as shown by the solid curve in
the left panel.
In contrast to the light curve produced only by eddies,
the amplitude of subjacent smooth component in the
light curve with contribution of both eddies and inter-eddies is significantly
larger, as shown in the right panel.
Obviously, it looks more close to that of observations.
Now, we argue that the gamma-ray emission from inter-eddies
should be included in modeling the light curves of gamma-ray emission
in the relativistic turbulence model.

According to the numerical results, the amplitude of
subjacent smooth component may be $f$-dependent.
Figure~\ref{F:Ratio} shows the value of $\zeta$,
which is defined as the ratio of the fast variability amplitude to that of total gamma-ray emission,
as a function of $f$.
Here, the orientation of eddy velocity is isotropic in the shell frame,
and the value of $\zeta$ is calculated around the peak flux of total gamma-ray emission, i.e.,
$t\Gamma^2c/R_0 \sim [0.9, 1.1]$.
In order to suppress the fluctuations, we perform 40 simulations for
each situation.
This figure shows that the contribution of the fast variability to the total gamma-ray emission is almost proportional to $f$,
i.e., the amplitude of subjacent smooth component is large for low value of $f$.
Then, an appropriate value of $f$ is required for producing a light curve
close to the observation.

In the following part, we focus on the lead/lag of
gamma-ray emission with respect to the observed synchrotron emission,
which is completely dominated by that from inter-eddies (\citealp{Kumar2009}).
In this work, we concentrate on the effects of gamma-ray emission
from eddies on the lead/lag.
For simplicity, the lag of gamma-ray emission from inter-eddies
with respect to the observed synchrotron emission is ignored.
In other words, the light curve of gamma-ray emission from inter-eddies
is used to represent that of the observed synchrotron emission.
In addition, we would point out that an additional lead/lag, which may appear when a light curve is plotted in a given energy range rather than the total energy range of gamma-ray emission (or synchrotron emission), is also neglected in this work.
According to the above description,
the lead/lag will depend on the value of $f$, which describes the contribution of eddies to gamma-ray emission.
If $f \lesssim 0.5$, the gamma-ray emission is mainly from inter-eddies,
and thus the absolute value of lead/lag may be low.
This behavior can be found in Figure~\ref{F:Lag_f}.
In this figure,
the circles correspond to the numerical results with $|\mu'| \geqslant 0.8$
and the squares correspond to the numerical results with $|\mu'| \leqslant 1$.
The positive (negative) values of $t_{\rm lag}\Gamma^2 c /R_0$ indicate that
the gamma-ray emission leads (lags behind) the observed synchrotron emission.
As shown in the figure, the absolute value of
lead/lag between gamma-ray emission and synchrotron emission
decreases as $f$ approaches $0$.
Then, we choose two cases, i.e., $f=0.6$ and $f=0.8$, for our study on
the lead/lag behavior.

Figure~\ref{F:Lag_Mu} shows the lead/lag in different $|\mu'|$ conditions.
In this figure, the squares represent the numerical results with $|\mu'| \geqslant a$,
and the circles represent the numerical results with $|\mu'| \leqslant a$.
The empty symbols are for $f=0.8$, and the filled symbols are for $f=0.6$.
As shown in the figure, the lead/lag for different $|\mu'|$ conditions is quite different.
The simulations with $|\mu'|{\geqslant}a$ mainly produce
the lead of gamma-ray emission with respect to the observed synchrotron emission,
whereas the simulations with $|\mu'|{\leqslant}a$ produce the opposite behavior.
Obviously, the lead/lag in the simulations is related to the
distribution of eddy orientation rather than the radiation mechanism.
We therefore argue that the lead/lag found in non-stationary
short-lived light curves may not reveal the intrinsic lead/lag of
different emission components.

The physical reason for the lead/lag is related to
the difference between the peak time of gamma-ray emission from eddies and that from inter-eddies.
Figure~\ref{F:Lag_Schematic} gives two examples of gamma-ray emission
light curves to show the lead/lag behavior with $f=0.8$,
where the solid curves correspond to the gamma-ray emission
from eddies, and the dashed lines correspond to that from inter-eddies.
In this figure, the left panel is for $|\mu'|=1$, which is a typical example
for the situation with $|\mu'|{\geqslant}a$.
The right panel is for $\mu'=0$, which is a typical example for
the situation with $|\mu'|{\leqslant}a$.
For $|\mu'|=1$, the velocity of eddies is in the radial direction.
Since the gamma-ray emission from eddies with $\mu'=-1$ is negligible
compared with that from $\mu'=1$,
we use the situation of $\mu'=1$ to represent that of $|\mu'|=1$.
In this case, the gamma-ray emission of eddies can be viewed as the emission from
a jet with Lorentz factor of $2\Gamma\gamma'_t$ according to Equation (\ref{eq: Lorentz_Transform}).
In addition, the angular spreading time, which affects the peak time of gamma-ray emission,
is inversely proportional to the square of jet Lorentz factor.
Then, the gamma-ray emission from eddies will reach
its peak luminosity ahead of the emission from inter-eddies,
owing to the fact that Lorentz factor of jet is less than that of eddies.
This behavior is clearly shown in the left panel.
The corresponding result is that, the peak time of total gamma-ray emission
is ahead of the peak time of gamma-ray emission from inter-eddies,
and therefore ahead of the peak time of synchrotron emission.
On the other hand, for $\mu'=0$, the polar angle $\theta$ of eddies
is $\sim 1/\Gamma$ according to Equation~(\ref{eq: Angle_Transform}).
In this situation,
eddies in high latitude ($\theta_{\rm j}> 1/\Gamma$) of the jet will make
more contribution to the gamma-ray emission than that in low latitude.
The corresponding result is that
the peak time of gamma-ray emission from eddies is behind that
from inter-eddies, and therefore
the peak time of total gamma-ray emission is behind that of
synchrotron emission.
Thus, the simulation results in Figure~\ref{F:Lag_Mu}, either for $|\mu'|\geqslant a$
or for $|\mu'|\leqslant a$, can be well understood.

\section{Discussion and Conclusions}

In this work, we have studied the light curves of GRBs in
the relativistic turbulence model by considering the role of
inter-eddy emission.
By ignoring the lag between gamma-ray emission from inter-eddies and the observed synchrotron emission,
our numerical simulations for the light curves
show that the gamma-ray emission can either lead or lag behind the
observed synchrotron emission.
The lead/lag is due to the different peak time of
gamma-ray emission in different situation,
which is related to the angular spreading time.
We argue that the lead/lag found in non-stationary
short-lived light curves may not reveal the intrinsic lead/lag of
different emission components.

\begin{figure}
\plotone{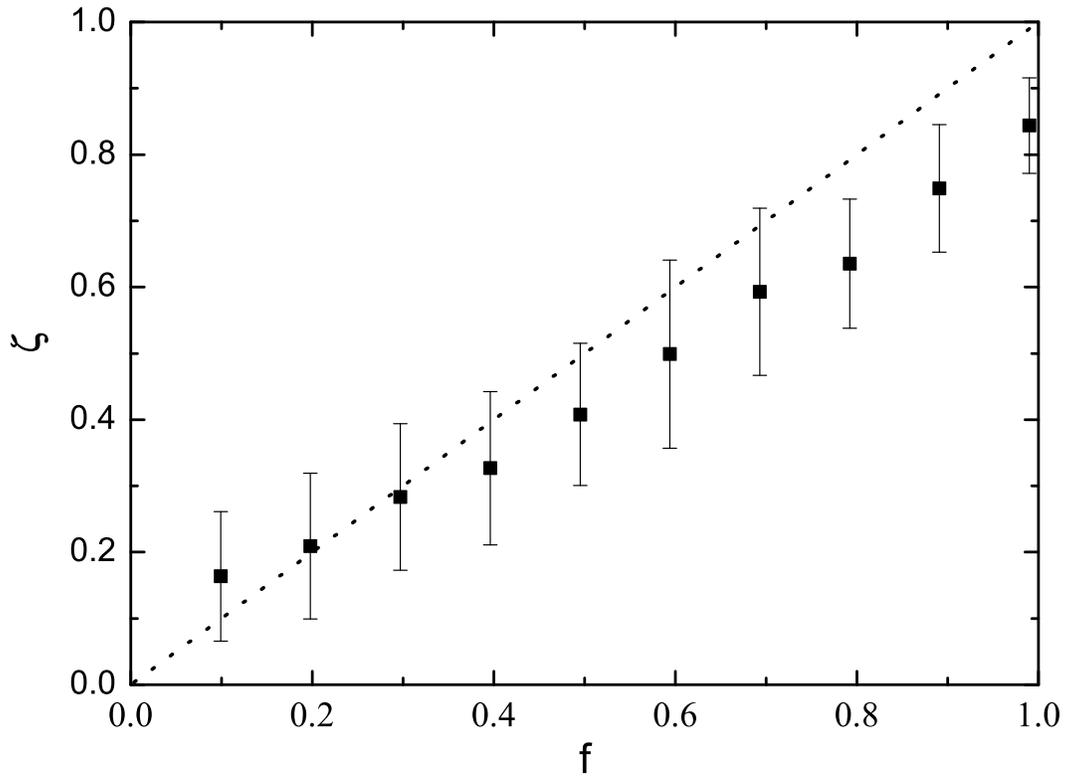}
\caption{Relationship between $\zeta$ and $f$ based on simulations.
The dotted line corresponds to $\zeta=f$.}
\label{F:Ratio}
\end{figure}

\begin{figure}
\plotone{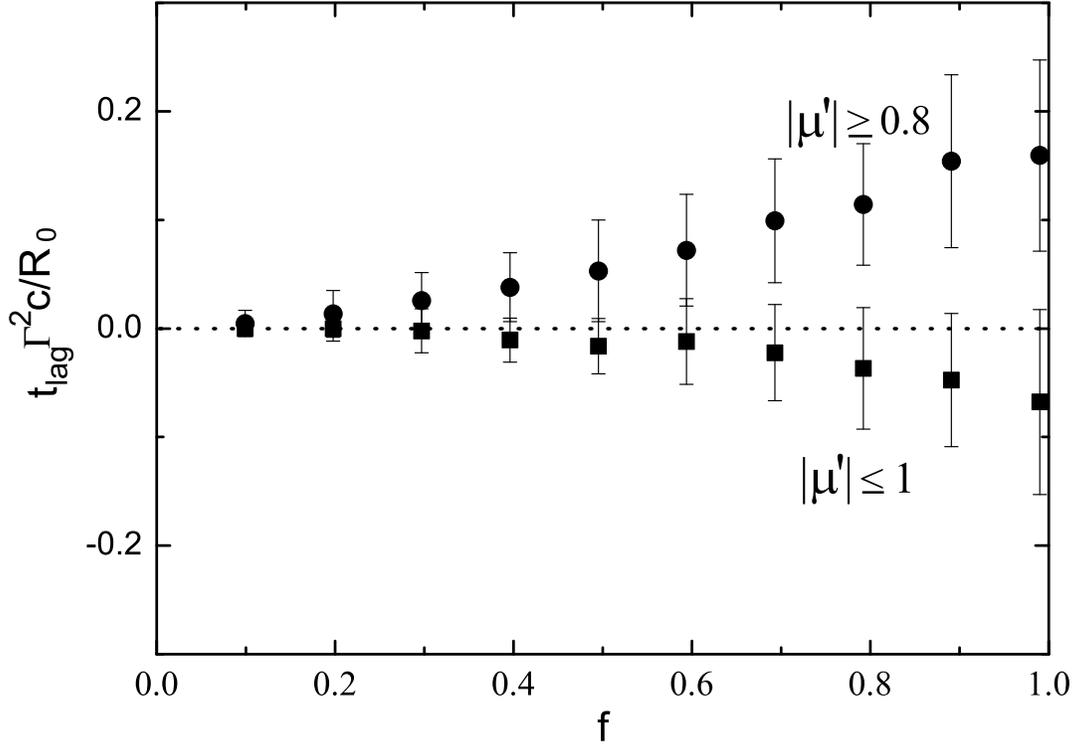}
\caption{Lead/lag of the gamma-ray emission with respect to
the observed synchrotron emission for various $f$,
where the circles correspond to $|\mu'| \geqslant 0.8$
and the squares correspond to $|\mu'| \leqslant 1$.
The positive (negative) vertical values indicate that
the gamma-ray emission leads (lags behind) the observed synchrotron emission.}
\label{F:Lag_f}
\end{figure}

\begin{figure}
\plotone{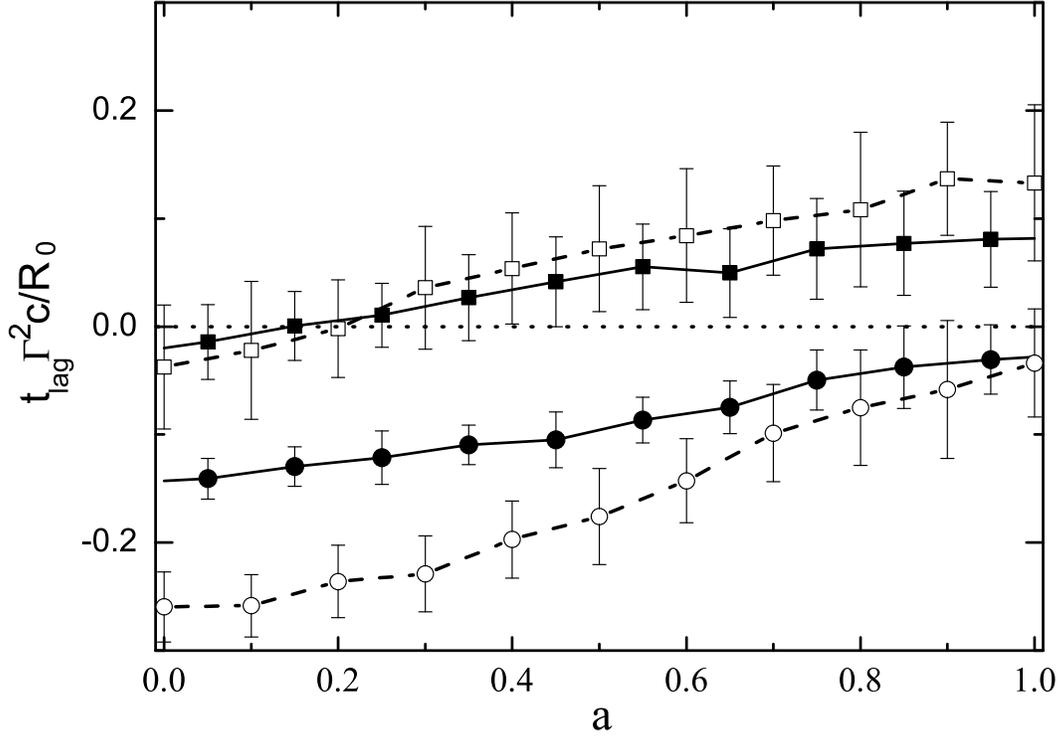}
\caption{Lead/lag of the gamma-ray emission with respect to
the observed synchrotron emission for different $|\mu'|$ conditions.
The empty symbols are for $f=0.8$ and the filled symbols are
for $f=0.6$.
The squares represent the numerical results with $|\mu'|\geqslant a$,
and the circles represent the results with $|\mu'|\leqslant a$.}
\label{F:Lag_Mu}
\end{figure}

\begin{figure}
\plotone{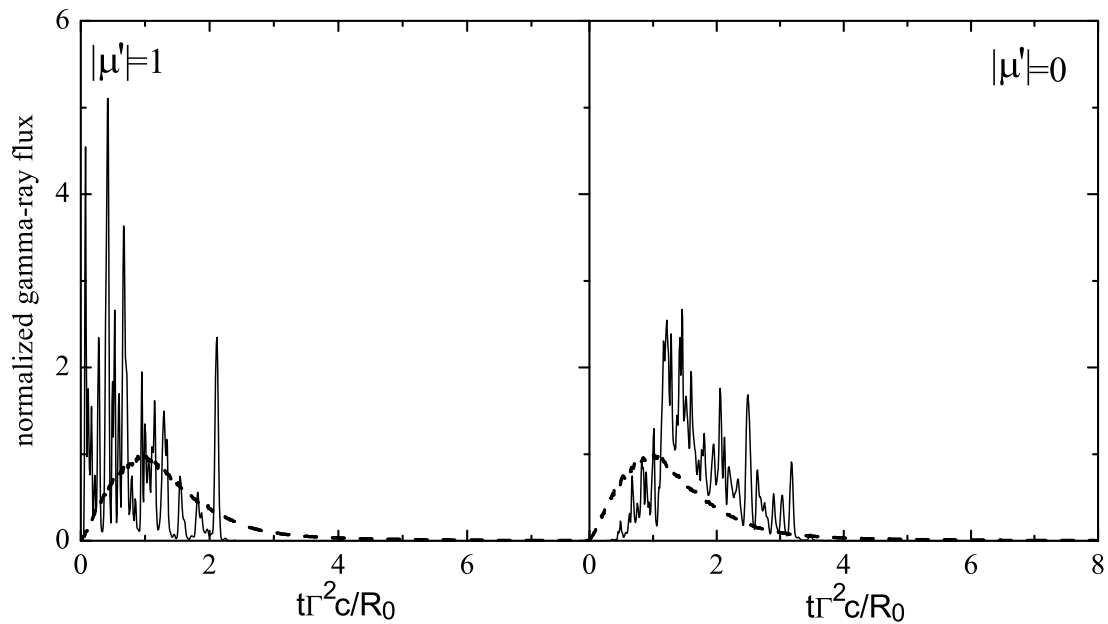}
\caption{Two examples of light curves with $f=0.8$,
where the solid curves correspond to the gamma-ray emission from eddies and
the dashed lines correspond to the gamma-ray emission from inter-eddies.
The left panel is for $|\mu'|=1$, which shows that the peak time of
gamma-ray emission from eddies is ahead of that from inter-eddies.
The right panel is for $|\mu'|=0$, which shows the opposite behavior.}
\label{F:Lag_Schematic}
\end{figure}

For GRB 080319B, Figure~4 of \citet{Wozniak2009} implies
$\zeta \sim 0.5-0.7$, which corresponds to the filling factor
$f \sim 0.6-0.8$ according to our Figure~\ref{F:Ratio}.
Since the duration of main episode in this burst is
$\Gamma ^2 c/R_0 \sim 28$~s (see \citealp{Patricelli2012} and references therein),
the lead/lag of gamma-ray emission with respect to optical
(synchrotron) emission is probably in the range of
$[-7.3\;{\rm s}, 3.7\;{\rm s}]$ based on our Figure~\ref{F:Lag_Mu}.
On the other hand, \citet{Beskin2010} showed that the lead of
gamma-ray emission to optical emission in this burst is around
$2~{\rm s}$, which is well in the above range.
Thus, such a lead may not rule out
the inverse-Compton scattering as a radiation mechanism
for producing the gamma-ray emission.

\acknowledgments
We thank the anonymous referee for helpful suggestions to improve the paper.
We also thank Bing Zhang for beneficial discussions.
This work was supported by the National Basic Research Program
(973 Program) of China under grant 2014CB845800,
the National Natural Science Foundation of China under grants
11073015, 11103015, 11222328, 11233006, 11025313, and
Guangxi Science Foundation under grant 2013GXNSFFA019001.

\clearpage

\clearpage

\clearpage

\clearpage

\clearpage

\end{document}